\documentclass[twocolumn]{jpsj3}
\usepackage{color}

\title{
Kondo effect in CeX$_{\text{c}}$ (X$_{\text{c}}$=S, Se, Te) studied by electrical resistivity under high pressure
}

\author{Yuya Hayashi$^{1}$, Shun Takai$^{1}$, Takeshi Matsumura$^{1,2}$, Hiroshi Tanida$^{1}$, Masafumi Sera$^{1,2}$, Kazuyuki Matsubayashi$^{3}$, Yoshiya Uwatoko$^{3}$, and Akira Ochiai$^{4}$ }

\inst{$^{1}$Department of Quantum Matter, AdSM, Hiroshima University, Higashi-Hiroshima, Hiroshima 739-8530, Japan \\
$^{2}$Institute for Advanced Materials Research, Hiroshima University, Higashi-Hiroshima, Hiroshima 739-8530, Japan \\
$^{3}$Institute for Solid State Physics, University of Tokyo, Kashiwa, Chiba, 277-8581, Japan \\
$^{4}$Department of Physics, Graduate School of Science, Tohoku University, Sendai 980-8578, Japan \\
}

\abst{
We have measured the electrical resistivity of cerium monochalcogenices, CeS, CeSe, and CeTe, under high pressures up to 8 GPa. 
Pressure dependences of the antiferromagnetic ordering temperature $T_{\text{N}}$, crystal field splitting, and the $\ln T$ anomaly of the Kondo effect have been studied to cover the whole region from the magnetic ordering regime at low pressure to the Fermi liquid regime at high pressure. 
$T_{\text{N}}$ initially increases with increasing pressure, and starts to decrease at high pressure as expected from the Doniach's diagram. Simultaneously, the $\ln T$ behavior in the resistivity is enhanced, indicating the enhancement of the Kondo effect by pressure. 
It is also characteristic in CeX$_{\text{c}}$ that the crystal field splitting rapidly decreases at a common rate of $-12.2$ K/GPa. 
This leads to the increase in the degeneracy of the $f$ state and further enhancement of the Kondo effect. 
It is shown that the pressure dependent degeneracy of the $f$ state is a key factor to understand the pressure dependence of $T_{\text{N}}$, Kondo effect, magnetoresistance, and the peak structure in the temperature dependence of resistivity. 
 }

\begin{document}
\setlength{\textwidth}{504pt}
\setlength{\columnsep}{14pt}
\hoffset-23.5pt
\setlength{\textheight}{680pt}

\maketitle
\section{Introduction}
Through the hybridization between conduction electrons and localized $f$ state ($c$-$f$ hybridization), the $f$ electrons at different atomic sites interact with each other, leading to a magnetic ordered state. This is the Ruderman-Kittel-Kasuya-Yosida (RKKY) interaction. 
At the same time, the $c$-$f$ hybridization causes the local spin and orbital degrees of freedom to be screened by the conduction electrons, leading to the reduction of the local moment and the magnetic ordering temperature. 
This is the Kondo effect. 
The strength of the $c$-$f$ hybridization, $J_{cf}$, is the primary parameter to describe these competing phenomena. 
In Ce based compounds with one $4f$ electron, the RKKY interaction dominates the Kondo effect when $J_{cf}$ is small, and the magnetic ordering temperature increases with increasing $J_{cf}$. 
With further increasing $J_{cf}$, the Kondo effect gradually dominates the RKKY interaction, and the magnetic ordering temperature starts to decrease. Finally, the magnetic ordering vanishes, and the $c$-$f$ hybridized system forms a Fermi liquid ground state with heavy effective mass. 
This general scheme has been described by Doniach's diagram.~\cite{Doniach77,Otsuki09} 
Of particular interest in recent years is in the vicinity of $J_{cf}$ where the magnetic ordering vanishes, which is called the quantum critical point (QCP).  
In this region, the order parameter fluctuation causes the physical properties to deviate from the Fermi liquid behavior, and in some cases causes unconventional superconductivity.\cite{Lohneysen07,Knebel01,Ren14} 
In the course of this study, application of pressure is an suitable method to experimentally realize the variation of $J_{cf}$.

CeS, CeSe, and CeTe, crystallizing in a cubic NaCl-type structure, exhibit antiferromagnetic (AFM) orderings with $\mib{q}=(\frac{1}{2}, \frac{1}{2}, \frac{1}{2})$ at $T_{\text{N}}$=8.4, 5.4, and 1.9 K, respectively.\cite{Hulliger78,Ott79,Schoenes87}
In a cubic crystalline electric field (CEF), the Hund's rule ground multiplet of $J=5/2$ splits into the $\Gamma_7$ doublet and the $\Gamma_8$ quartet. 
It is already well established that the ground state of CeX$_{\text{c}}$ is the $\Gamma_7$ and the energy level of the $\Gamma_8$ state is at 140 K, 116 K, and 32 K for CeS, CeSe, and CeTe, respectively.\cite{Rossat-Mignod85,Donni93a,Donni93b} 
The magnitude of the ordered moment in the AFM phase is 0.57 $\mu_{\text{B}}$, 0.56 $\mu_{\text{B}}$, and 0.3 $\mu_{\text{B}}$ for CeS, CeSe, and CeTe, respectively. 
The reduction of the ordered moment from the expected value of 0.71 $\mu_{\text{B}}$ for the $\Gamma_7$ ground state can be ascribed to the Kondo effect. 
However, the largest reduction in CeTe cannot be understood from the Kondo effect only, because the Kondo effect  is the weakest in CeTe and the strongest in CeS as judged from the $\ln T$ anomaly ($d\rho/d\ln T$) in resistivity.\cite{Schoenes87} 
This is still an open question.

In most of the Ce based compounds that have been studied to date under high pressure in the quantum critical region, the CEF ground states are well isolated Kramers doublets, where only the magnetic dipolar degree of freedom plays a role. 
It is also the case in CeX$_{\text{c}}$ at ambient pressure with the $\Gamma_7$ ground state. 
However, it has been strongly suggested by recent experiments on CeTe that the $\Gamma_8$ excited level falls down under high pressure.\cite{Kawarasaki11,Takaguchi15} 
The CEF splitting is expected to vanish at around 2.5 GPa in CeTe. 
This means that the quadrupolar degree of freedom of the $\Gamma_8$ state also takes part in the physical properties at low temperatures.
It has already been shown for CeTe that an antiferroquadrupolar interaction through the $\Gamma_8$ excited level is essential to understand the properties under high pressure.\cite{Takaguchi15} 
Therefore, nontrivial phenomenon could take place in the quantum critical region by an interplay between nondipolar degrees of freedom and the Kondo effect. 
This is also an important issue in relation to the recent studies in Pr based compounds, where the role of quadrupolar fluctuation for the superconductivity attracts interest.\cite{Matsubayashi12} 
High pressure study on CeX$_{\text{c}}$, therefore, has its significance in studying the possibility of multipolar degrees of freedom to participate in the quantum critical phenomenon. 

Another advantage in studying CeX$_{\text{c}}$ lies in its simple electronic structure. 
The localized $f$ state hybridizes with the $p$-orbital of X$_{c}$, which forms the valence band. The conduction band is formed by the $5d$-orbital of Ce, containing one conduction electron per formula unit. The $f$ level is located $\sim2.3$ eV below the Fermi level. 
These features common in CeX$_{\text{c}}$ have been clarified in detail by the angle-resolved photoemission spectroscopy and the de Haas-van Alphen effect measurements.\cite{Chiaia98,Nakayama04a,Nakayama04b}

In the present work, we have measured the electrical resistivity of CeX$_{\text{c}}$ under high pressures up to 8 GPa to study the enhancement of the Kondo effect and the crossover from the magnetic ordering regime to the Fermi liquid regime. 
We show that the $\ln T$ behavior in the resistivity is enhanced by applying pressure in all the CeX$_{\text{c}}$ compounds. 
The pressure dependences of $T_{\text{N}}$, CEF splitting, coefficient of the $\ln T$ term, 
and the temperatures of resistivity maximum below which the Fermi liquid state is formed, are extracted from the data. 
Although the overall feature can be understood in the framework of Doniach's diagram, 
we suggest that the level lowering of the $\Gamma_8$ excited state at high pressures gives rise to some distinct phenomena in CeX$_{\text{c}}$ in the crossover region around the QCP. We compare these results with those of other inter-metallic compounds.

\section{Experiment}
The single crystalline samples were prepared by Bridgman method as described in Ref.~\citen{Nakayama04b}. 
Resistivity under high pressure was measured by normal four terminal method. The current direction was along the [001] axis. 
For measurements up to 2.5 GPa, we used a CuBe/NiCrAl hybrid-piston-cylinder cell with daphne oil as a pressure transmitting medium.
Magnetic field was also applied using a 15 T cryomagnet. The field direction was along the [001] axis, which was parallel to the current direction. 
The measurements up to 8 GPa were performed at zero field using a cubic anvil cell with fluorinert as a pressure transmitting medium.

\section{Experimental Results}
\subsection{Resistivity at zero field}
\begin{fullfigure}[t]
\begin{center}
\includegraphics[width=17cm]{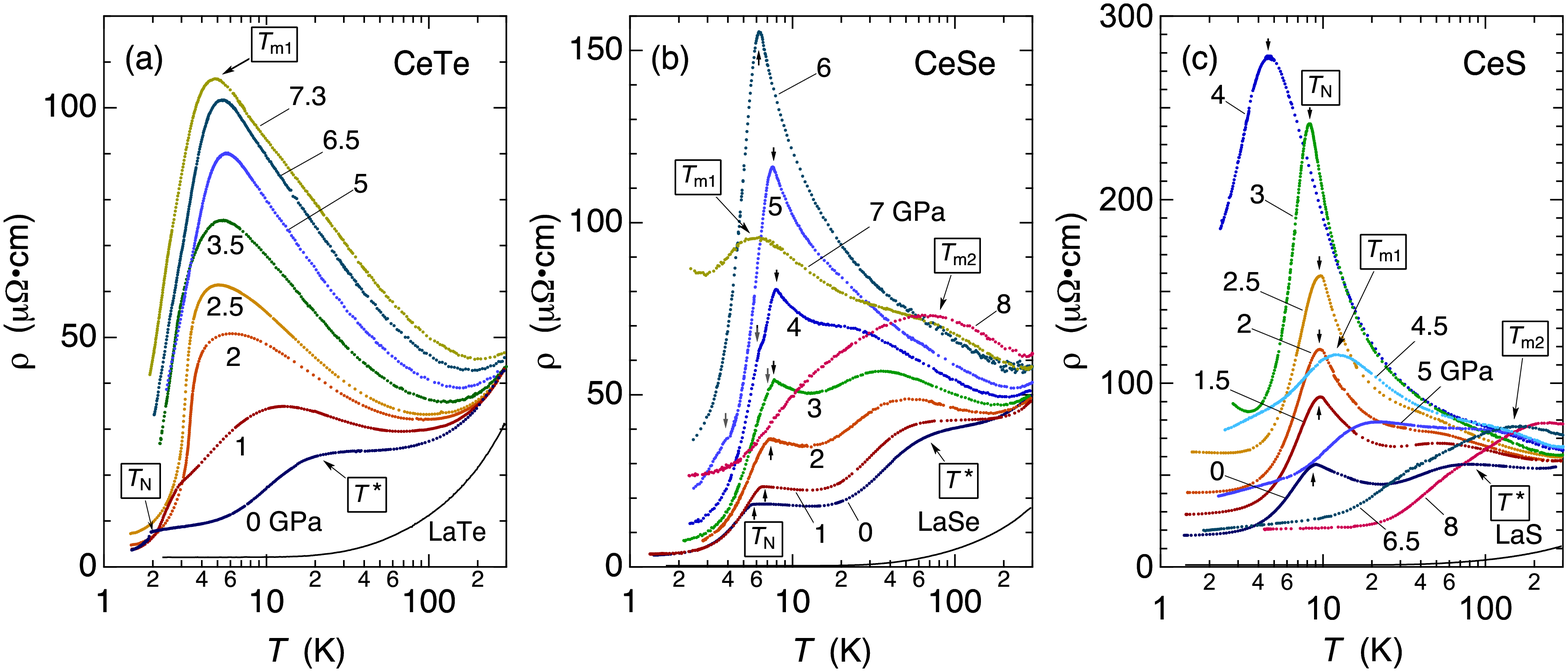}
\end{center}
\caption{(Color online) Temperature dependences of the electrical resistivity of (a) CeTe, (b) CeSe, and (c) CeS, under high pressures. 
The resistivity of LaTe, LaSe, and LaS, at ambient pressure, are also shown for reference. }
\label{fig:1}
\end{fullfigure}

\begin{fullfigure}[t]
\begin{center}
\includegraphics[width=17cm]{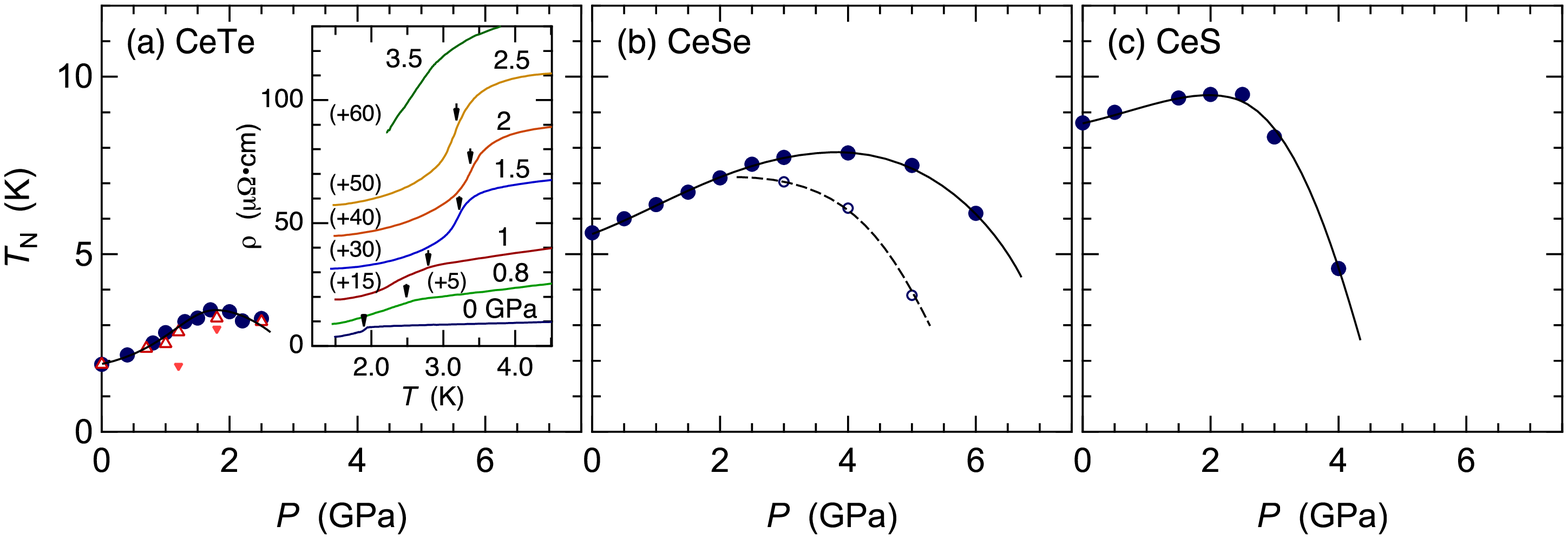}
\end{center}
\caption{(Color online) Pressure dependences of $T_{\text{N}}$ of (a) CeTe, (b) CeSe, and (c) CeS, determined from $\rho(T)$ (open circles).  The inset of (a) shows $\rho(T)$ of CeTe in the vicinity of $T_{\text{N}}$. 
Open and filled triangles in (a) represent the transition temperatures of CeTe determined from the specific heat measurement in Ref.~\citen{Takaguchi15}. The filled triangles at 1.2 GPa and 1.8 GPa show the successive transitions. 
Open circles in (b) correspond to the weak anomalies observed in $\rho(T)$ of CeSe below $T_{\text{N}}$. }
\label{fig:2}
\end{fullfigure}

\subsubsection{CeTe}
Figure~\ref{fig:1} (a) shows the temperature ($T$) dependences of the resistivity ($\rho$) of CeTe at several pressures. 
The $\rho(T)$ of LaTe is also shown for reference to estimate the contribution from phonon scattering. 
At ambient pressure, a hump anomaly is observed around 20 K, reflecting the CEF splitting; $\rho$ decreases below $\sim 20$ K because the magnetic scattering through the $\Gamma_8$ exited state gradually vanishes. 
We name this characteristic temperature $T^*$. 
With further decreasing temperature, $\rho$ drops steeply at 2 K, reflecting the AFM order. 
The $\ln T$ behavior in $\rho(T)$ due to the Kondo effect can be recognized above $T^*$ after subtracting the $\rho(T)$ of LaTe, 
whereas it is not recognizable below $T^*$. 
These behaviors are consistent with the previous study.\cite{Schoenes87}

With increasing $P$, $\rho(T)$ exhibits a clear $\ln T$ dependence below $\sim 100$ K even before subtracting the $\rho(T)$ of LaTe. 
This shows that the Kondo effect (single site Kondo scattering) is enhanced by the pressure. 
If we subtract the $\rho(T)$ of LaTe, the $\ln T$ dependence continues up to 300 K. 
The hump anomaly at $T^*$ shifts to lower temperatures, corresponding to the level lowering of the $\Gamma_8$ exited state.\cite{Kawarasaki11}
However, above 2.5 GPa, the hump anomaly is no more distinguishable from the peak emerging around 6 K, probably reflecting the Kondo coherence. We name this temperature of resistivity maximum as $T_{\text{m1}}$ hereafter. 
When $P$ is increased from 3.5 GPa to 5 GPa, the coherence peak slightly shifts to higher temperature, which seems to reflect the increase in the Kondo temperature $T_{\text{K}}$. 
However, the peak strangely shifts again to lower temperature at 7.3 GPa, which is opposite to the normal $P$ dependence of the coherence peak in Ce based compounds. 
Although this could be somehow related with the transition to the CsCl structure at 8 GPa in CeTe,\cite{Leger83} the reason is unclear. 

The low temperature part of $\rho(T)$ around $T_{\text{N}}$ is shown in the inset of Fig.~\ref{fig:2} (a). 
As indicated by the arrows, $T_{\text{N}}$ of CeTe initially increases with increasing $P$, and then starts to decrease above $\sim 2$ GPa. At 3.5 GPa, the anomaly in $\rho(T)$ corresponding to $T_{\text{N}}$ is no more able to be identified. 
The $P$ dependence of $T_{\text{N}}$ is plotted in Fig.~\ref{fig:2} (a). 
$T_{\text{N}}$ determined from the specific heat measurement in Ref.~\citen{Takaguchi15} is also plotted. 
With respect to the successive transitions at 1.2 GPa and 1.8 GPa reported in Ref.~\citen{Takaguchi15}, the higher transition temperature agree with $T_{\text{N}}$ deduced from $\rho(T)$. 
Anomaly corresponding to the second transition at a lower temperature, which is indicated by the filled triangles in Fig. ~\ref{fig:2} (a), was not detected in $\rho(T)$ as a clear anomaly. 
The critical pressure, $P_{\text{c}}$, is estimated to be from 3 to 3.5 GPa, although this requires a further study at lower temperatures.

\subsubsection{CeSe}
Figure 1 (b) shows the $\rho(T)$ curves of CeSe up to 8 GPa. 
The $\rho(T)$ of LaSe is also shown for reference to estimate the contribution from phonon scattering. 
At ambient pressure, in the same way as in CeTe, the hump anomaly reflecting the CEF splitting appears at around $T^*=60$ K and the kink anomaly due to the AFM order is clearly observed at 5.4 K. 
The $\rho(T)$ weakly increases with decreasing $T$ below $\sim 20$ K down to $T_{\text{N}}$. 
This increase can be ascribed to the Kondo effect within the $\Gamma_7$ ground state. 
The $\ln T$ behavior in the high-$T$ region above $T^*$ can be recognized after subtracting the $\rho(T)$ of LaSe. 
These behaviors at ambient pressure are consistent with the previous study.\cite{Schoenes87} 

With increasing $P$, $\rho(T)$ increases and exhibits a clear $\ln T$ anomaly below $\sim 200$ K, which shows that the Kondo effect (single site Kondo scattering) is enhanced by the pressure. If we subtract the $\rho(T)$ of LaSe, the $\ln T$ dependence continues up to 300 K. 
Simultaneously, the hump anomaly around $T^*$ shifts to lower temperatures, indicating that the energy level of the $\Gamma_8$ exited state falls down with increasing $P$. 
Above 5 GPa, $T^*$ is no more recognizable because it is hidden behind the significant increase of $\rho(T)$ due to the Kondo effect. 

The $\ln T$ slope below $\sim 20$ K down to $T_{\text{N}}$, where the Kondo effect within the $\Gamma_7$ ground state dominates, also increases with $P$. 
Above 5 GPa, it is more enhanced and $\rho(T)$ exhibits a steep increase down to $T_{\text{N}}$, where a sharp cusp is observed. 
These behaviors are in contrast to CeTe at high pressures, where $\rho(T)$ follows a single $\ln T$ slope down to $T_{\text{m1}}\sim 6$ K and the cusp anomaly at $T_{\text{N}}$ soon becomes unclear at high pressures above 2 GPa. 
%The steep increase in $\rho(T)$ below $\sim 20$ K could also be associated with the magnetic scattering in the critical region near $T_{\text{N}}$, suggesting the existence of a large magnetic fluctuation. 

In CeSe, as indicated by the arrows in Fig. 1 (b), $T_{\text{N}}$ initially increases with $P$, taking a maximum at $\sim 4$ GPa, and decreases. 
The $P$ dependence of $T_{\text{N}}$ is plotted in Fig. 2 (b). 
The critical pressure, $P_{\text{c}}$, is estimated to be around 7.5 GPa. 
Above 3 GPa, we can also see another weak anomaly in $\rho(T)$ below the main cusp anomaly at $T_{\text{N}}$, 
which is plotted by the open circles. 
We speculate that this transition reflects some change in the magnetic structure such as the one we discussed in CeTe by a mean-field calculation.~\cite{Takaguchi15} 
Namely, in the intermediate phase, the AFM moment is possibly oriented along the [100] axis because of the $\Gamma_8$ level lowering, and in the low-$T$ phase, the direction changes to the [111] axis, reflecting the easy axis of the $\Gamma_7$ ground state. 

All these behaviors change significantly above 7 GPa, where $\rho(T)$ at low temperatures decreases significantly. 
At 7 GPa, a new broad peak appears at around 6 K, which we suggest the peak of $T_{\text{m1}}$, the same kind of Kondo coherence peak as the one observed in CeTe. 
The upturn in $\rho(T)$ below 3 K probably suggests the AFM ordering existing below the lowest temperature of 2.5 K. 
It is also remarked that there is a broad hump anomaly at around 60 K, which we name $T_{\text{m2}}$. 
At 8 GPa, the position of this peak shifts to around 70 K, above which the $\ln T$ behavior is still observed. 
At low temperatures below 70 K, $\rho(T)$ shows a more significant decrease down to the lowest temperature, suggesting the formation of a Fermi liquid state. 
Therefore, the broad peak at $T_{\text{m2}}$, 60 K at 7 GPa and 70 K at 8 GPa, may also be assigned to another Kondo coherence peak. 
The coherence peak at $T_{\text{m1}}$ still seems to exist at around 15 K, where a broad hump is observed. 
Thus, the double peak structure in $\rho(T)$ is the characteristic of CeSe in the Fermi liquid regime, and both $T_{\text{m1}}$ and $T_{\text{m2}}$ shift to higher temperatures with increasing $P$. 
%One possible origin of this peak is the CEF splitting because it lies on the extrapolation of the CEF anomaly at low pressures. 
%This behavior cannot be understood from the CEF level lowering effect. 

\subsubsection{CeS}
Figure 1 (c) shows the $\rho(T)$ curves of CeS at high pressures. 
At ambient pressure, in the same way as in CeTe and CeSe, the hump anomaly reflecting the CEF splitting appears at around $T^*=70$ K and the cusp anomaly due to the AFM order is clearly observed at 8.7 K. 
An increase in $\rho(T)$ with decreasing $T$ is observed below $\sim 20$ K down to $T_{\text{N}}$, which can be ascribed to the Kondo effect within the $\Gamma_7$ ground state. The slope of the $\ln T$ dependence in the low-$T$ region is larger than that of CeSe. 
The $\ln T$ dependence is also visible in the high-$T$ region above $T^*$ even before subtracting the $\rho(T)$ of LaS. 
The largest slopes of the $\ln T$ dependences in CeS at ambient pressure, both in the high- and low-$T$ regions, show that the $c$-$f$ hybridization is the strongest among the three compounds of CeX$_{\text{c}}$. 
These behaviors at ambient pressure are consistent with the previous study.\cite{Schoenes87} 

With increasing $P$, the $\ln T$ slope increases as in CeTe and CeSe, reflecting the enhancement of the Kondo effect (single site Kondo scattering). 
%It is noted, however, the increasing rate of the $\ln T$ slope is in the same order as in CeTe and CeSe only at high temperatures above $T^*$ in CeS. 
A drastic phenomenon in CeS is that $\rho(T)$ in the low-$T$ region increases more steeply than the $\ln T$ slope in the high-$T$ region. This behavior becomes more significant above 2.5 GPa. 
The increase in $\rho(T)$ on approaching $T_{\text{N}}$, which can probably be ascribed to the enhancement of the Kondo effect, is much larger than that of CeSe. 
With respect to $T_{\text{N}}$, it initially increases with increasing $P$, followed by a maximum around 2 GPa, and decreases to 4.5 K at 4 GPa. 
The upturn in $\rho(T)$ of 3 GPa below 3.5 K probably suggests another AFM ordering existing below the lowest temperature of 2.8 K. 
The $P$ dependence of $T_{\text{N}}$ is plotted in Fig. 2 (c). 
The critical pressure, $P_{\text{c}}$, is estimated to be around 4.5 GPa.

Above 4.5 GPa, a significant change takes place at low temperatures. 
Although the $\ln T$ behavior remains at high temperatures above 100 K, $\rho(T)$ at low temperatures decreases significantly with increasing $P$. 
At 4.5 GPa, a new broad peak appears at around 12 K, which we suggest the peak of $T_{\text{m1}}$, the same kind of Kondo coherence peak as the one observed in CeTe and in CeSe.  
It is again remarked that there is another broad hump anomaly at around 100 K, which we name $T_{\text{m2}}$. 
With further increasing $P$, both of these broad peaks shift to higher temperatures. 
Finally at 8 GPa, the $\rho(T)$ curve becomes like that of a typical Fermi liquid system. 
However, the two broad peaks at $T_{\text{m1}}$ and $T_{\text{m2}}$ still remain in $\rho(T)$ at 8 GPa; one is at around 60 K and the other is at around 200 K, both of which may be ascribed to the coherence peak in the same way as in CeSe.

\subsection{Magnetoresistance}
\begin{fullfigure}[t]
\begin{center}
\includegraphics[width=15cm]{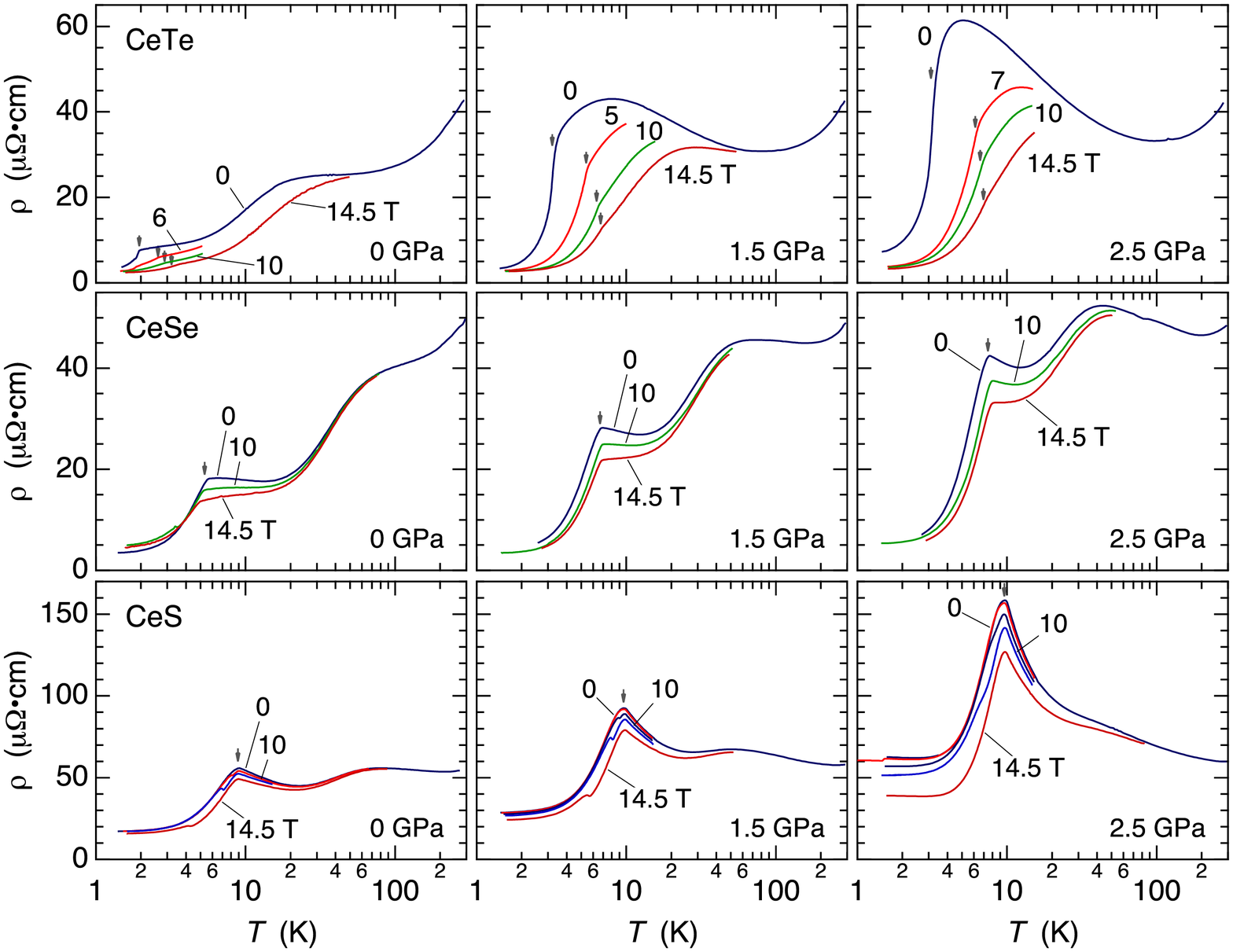}
\end{center}
\caption{(Color online) Temperature dependences of electrical resistivity in magnetic fields and under high pressures for CeTe, CeSe, and CeS. Arrows indicate the temperatures of antiferromagnetic ordering. 
}
\label{fig:3}
\end{fullfigure}
\begin{fullfigure}[t]
\begin{center}
\includegraphics[width=15cm]{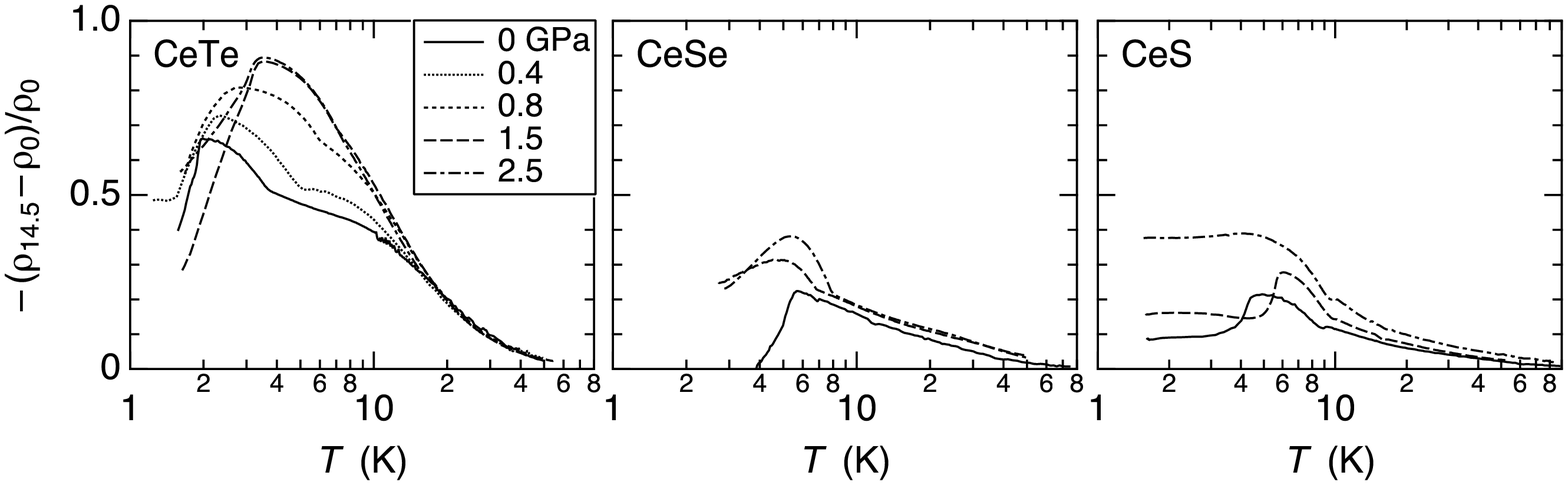}
\end{center}
\caption{(Color online) Temperature dependences of the magnetoresistance at 14.5 T, $-(\rho_{14.5} - \rho_{0})/\rho_{0}$, at several pressures. 
}
\label{fig:4}
\end{fullfigure}

Figure \ref{fig:3} shows the temperature dependences of electrical resistivity in magnetic fields and under high pressures for CeTe, CeSe, and CeS. All the compounds generally exhibit negative magnetoresistance (MR). 
In CeTe, the MR effect increases significantly with increasing $P$. 
At 1.5 GPa and 2.5 GPa, where the Kondo effect is clearly recognized in the $\ln T$ behavior in $\rho(T)$ at zero field, 
the $\rho$ value decreases and the peak in $\rho(T)$ at $T_{\text{m1}}$, 7 K and 5 K, respectively, shifts to higher temperatures with increasing the field. 
This is a characteristic behavior observed in concentrated Kondo systems.\cite{Takase80,Kawakami86} 
In CeSe and CeS, by contrast, although the MR effect increases with increasing $P$, it is not so significant in comparison with that of CeTe. 
This is because the system is still well inside the magnetic ordering regime at these pressures below 2.5 GPa. 
The Kondo temperature $T_{\text{K}}$ for the $\Gamma_7$ ground state may be much smaller than $T_{\text{N}}$. 
In CeTe, on the other hand, $T_{\text{K}}$ is expected to be already larger than $T_{\text{N}}$ at pressures of 1.5 GPa and 2.5 GPa because of the high degeneracy of the ground state due to the almost collapsed CEF.\cite{Yamada84,Hanzawa85} 

To describe the MR associated with the Kondo effect in more detail, we show in Fig.~\ref{fig:4} the temperature dependences of the relative MR at 14.5 T for several pressures. 
We focus on the temperature range well above $T_{\text{N}}$: $T>5$ K for CeTe and $T>10$ K for CeSe and CeS. 
The magnitudes of MR in CeSe and CeS are almost the same. 
If we observe in detail, however, MR is almost independent of $P$ below 2.5 GPa in CeSe, whereas in CeS MR increases slightly with increasing $P$. 
 
It is noted that CeTe has the largest MR among the three compounds in spite of the fact that $J_{cf}$ is the smallest. 
This may be because the $f$-state degeneracy influenced by the low-lying $\Gamma_8$ excited level, which increases $T_{\text{K}}$, is the largest among the three compounds even at ambient pressure. 
With increasing $P$, the MR in CeTe increases below $\sim 15$ K, and it becomes almost independent of $P$ above 1.5 GPa. 
It is also remarked that the MR does not change with $P$ above $\sim 15$ K.

Finally, we comment on the magnetic ordering. 
In CeTe, $T_{\text{N}}$ exhibit a significant dependence both on the applied field and pressure, which can be explained by considering the magnetic and quadrupolar interaction in addition to the $\Gamma_8$ level lowering as we have shown in the previous study.~\cite{Kawarasaki11,Takaguchi15} 
In CeSe and CeS, by contrast, $T_{\text{N}}$ is little affected by the field in this $P$ range below 2.5 GPa. 
This is because the energy scale of the magnetic interaction is larger than that of CeTe, resulting in larger $T_{\text{N}}$, and also because the $\Gamma_8$ level is still located at high energy and therefore the level mixing does not affect the magnetic ordering. 
Finally, although this is not a subject of this work, weak anomalies observed in CeS below $T_{\text{N}}$ in Fig.~\ref{fig:3}, which are clearly identified at 0 GPa and 1.5 GPa in magnetic fields, are those reflecting the change in the magnetic structure or the domain distribution.\cite{Donni93a} 
We can see this phase boundary also in CeSe at 3.5 K in the data at 0 GPa and 10 T. 
This anomaly moves to $H<10$ T at 1.5 GPa and 2.5 GPa, which are not shown in Fig.~\ref{fig:3}.

\begin{fullfigure}[t]
\begin{center}
\includegraphics[width=15cm]{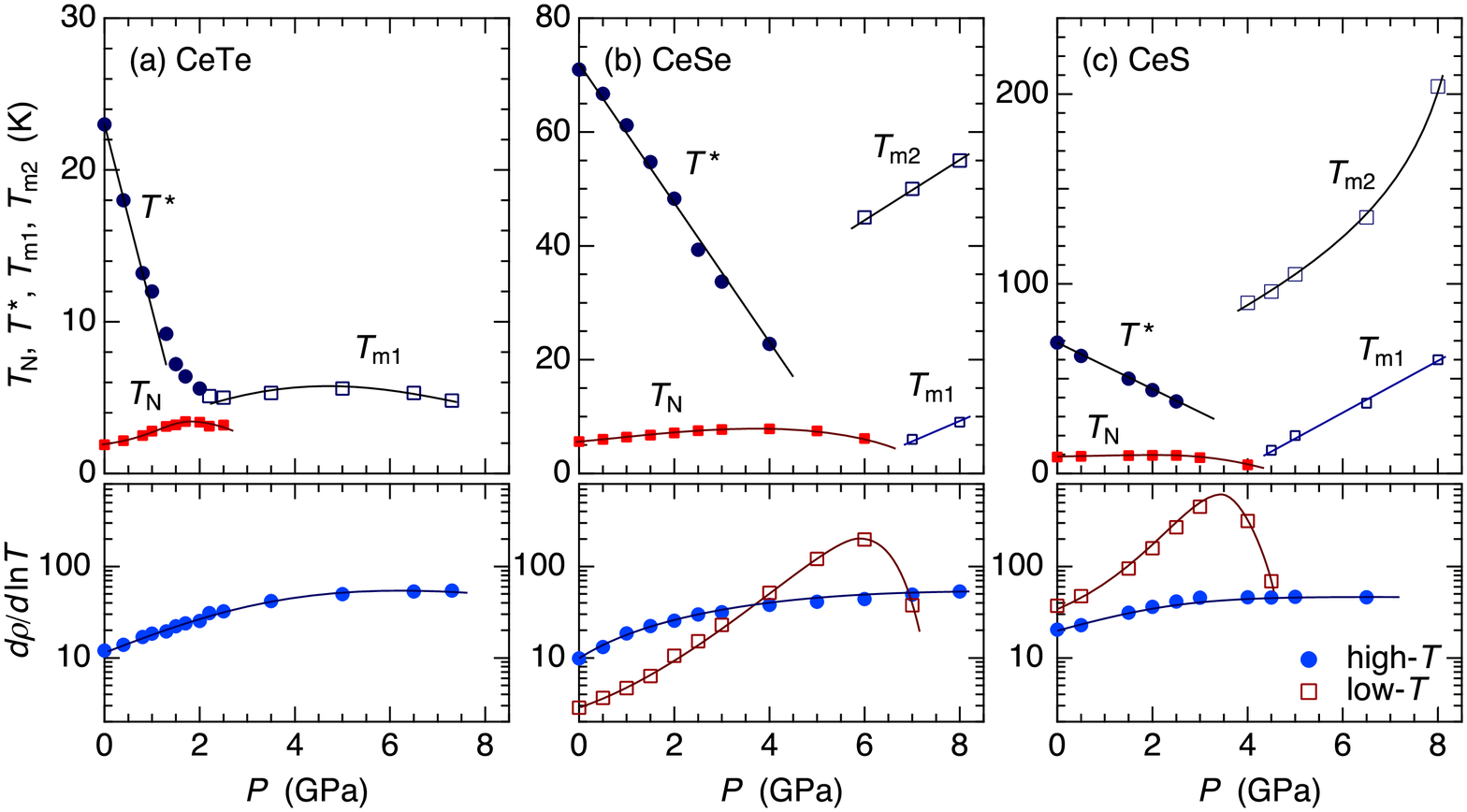}
\end{center}
\caption{(Color online) (Top panels) Pressure dependences of the characteristic temperatures $T_{\text{N}}$, $T^{*}$, $T_{\text{m1}}$, and $T_{\text{m2}}$. The lines are guides for the eye. It is remarked that all the three lines for $T^*$ are expressed as $T^*(P)=T^*(0)-12.2P$. 
(Bottom panels) Pressure dependences of the coefficient of the $\ln T$ term in the magnetic part of the resistivity, 
$d\rho_{\text{mag}}/d \ln T$, which are estimated in the high-$T$ region above $T^{*}$ or $T_{\text{m2}}$ and in the low-$T$ region between $T_{\text{N}}$ and $T^{*}$ or between $T_{\text{m1}}$ and $T_{\text{m2}}$. 
}
\label{fig:5}
\end{fullfigure}

\section{Discussion}

In Ce compounds, the coefficient of the $\ln T$ term in the magnetic part of the resistivity, $d\rho_{\text{mag}}/d \ln T \equiv S$, is generally associated with $J_{cf}D(\varepsilon_{F})$, where $D(\varepsilon_{F})$ represents the density of states of the conduction electrons at the Fermi energy. 
When the system is in the magnetic ordering regime in the Doniach's diagram, as in the present case of CeX$_{\text{c}}$, the coefficient $S$ and the $\rho$ value at low temperatures generally increase with increasing $P$. 
In addition, in this regime, a double peak structure is observed in $\rho_{\text{mag}}(T)$ in many systems. 
The examples are CeAl$_2$, CeCu$_2$, CeAu$_2$Si$_2$, CeNiGe$_3$, Ce$_2$Pd$_3$Si$_5$, and Ce$_2$Ni$_3$Ge$_5$.\cite{Ren14,Miyagawa08,Vargoz97,Link97,Nakashima04,Kurita08,Nakashima05} 
The peak at high temperature is generally interpreted as reflecting the CEF level splitting of the $f$ state and the second peak at low temperature as the formation of a coherent Kondo state. 
When the system is near the boundary between the magnetic ordering regime and the Fermi liquid regime, 
the $\rho_{\text{mag}}(T)$ generally exhibit a single peak at around the Kondo coherence temperature, and the peak shifts to high temperatures with increasing $P$ without changing the peak height. The peak position can finally reach to 300 K. 
The peak of CEF seems to be merged with the coherence peak. 
The examples are CeIn$_3$, CeAl$_3$ and CeCu$_6$.\cite{Knebel01,Kagayama94,Oomi96,Raymond00} 
To discuss the pressure dependence of these characteristic temperatures of CeX$_{\text{c}}$, in Fig.~\ref{fig:5} we plot $T_{\text{N}}$, $T^*$, $T_{\text{m1}}$, $T_{\text{m2}}$, and the slope $S$, which have been deduced from $\rho_{\text{mag}}(T)$, as a function of pressure. 
$\rho_{\text{mag}}(T)$ was obtained by subtracting $\rho(T)$ of LaX$_{\text{c}}$. 
We discuss these parameters in the following. 
The most important and distinctive feature we consider in CeX$_{\text{c}}$ is the shrinking of the CEF under pressure, which is reflected in the $P$ dependence of $T^*$. This effect leads to the increase in the $f$-state degeneracy and, therefore, the increase in $T_{\text{K}}$.\cite{Yamada84,Hanzawa85}   

First, we discuss the $P$ dependence of the $\ln T$ term in $\rho(T)$. 
When the CEF anomaly around $T^*$ is at high enough temperatures, we can extract two different coefficients of the $\ln T$ term from high-$T$ and low-$T$ regions above and below $T^*$, which we represent as $S_{\text{H}}$ and $S_{\text{L}}$, respectively. 
Pressure dependences of these coefficients are also shown in the bottom panels of Fig~\ref{fig:5}. 
In CeTe, since $T^*$ is soon merged into a peak at $T_{\text{m1}}$, only $S_{\text{H}}$ is shown. 
In CeSe and CeS, the two regions are clearly separated up to the critical pressure $P_{\text{c}}$ where $T_{\text{N}}$ vanishes. 
With respect to the pressure dependence of $S_{\text{H}}$, they almost coincide with each other at high pressures, although $S_{\text{H}}$ of CeS is slightly larger than others at $P=0$. 
We also see that the increasing rate of $S_{\text{H}}$ with pressure is almost the same in CeX$_{\text{c}}$.  
On the other hand, $S_{\text{L}}$ for CeS is almost ten times larger than that of CeSe and $S_{\text{L}}$ for CeTe is unable to recognize, 
which is probably because the Kondo scattering in CeTe is too small and is hidden behind the magnetic scattering due to disordered moments. 
This difference in $S_{\text{L}}$ shows that the Kondo effect in the low-$T$ region within the $\Gamma_7$ ground state is the strongest in CeS and the weakest in CeTe. 
By contrast, the similar values of $S_{\text{H}}$ seems to suggest that the Kondo effect in the high-$T$ region is more influenced by the $f$-state degeneracy involving all the CEF states than the $c$-$f$ hybridization effect. 

Another point to be noted is that $S_{\text{L}}$ exceeds $S_{\text{H}}$ in CeSe and in CeS at high pressures. 
Theoretically, the $\ln T$ coefficient is associated with the degeneracy, $\lambda$, of the $f$ state and is proportional to $\lambda^2 -1$.\cite{Cornut72} 
In the high-$T$ region, $\lambda=6$, and in the low-$T$ region when only the $\Gamma_7$ ground state is thermally populated, $\lambda=2$. Therefore, $S_{\text{H}}$ is generally larger than $S_{\text{L}}$, which is actually the case in many other inter-metallic compounds.\cite{Ren14,Miyagawa08}
In this sense, CeSe at low pressures is normal. 
Although the larger increasing rate of $dS_{\text{L}}/dP$ than $dS_{\text{H}}/dP$ is also observed in CeAl$_2$,\cite{Miyagawa08} 
it is particularly anomalous in CeSe and CeS that $S_{\text{L}}$ exceeds $S_{\text{H}}$. 

Next, with respect to $T_{\text{N}}$, $T^*$, $T_{\text{m1}}$, and $T_{\text{m2}}$, a general behavior observed in other inter-metallic compounds in the magnetic ordering regime can be summarized as the following (see Fig. 3 of Ref.~\citen{Ren14} or Fig. 2 of Ref.~\citen{Miyagawa08}): 
\begin{enumerate}
\item Double peaks appear at $T^*$ and $T_{\text{m1}}$ ($T^* > T_{\text{m1}}$). 
With increasing $P$, $T^*$ slightly decreases and $T_{\text{m1}}$ slightly increases. \\
\item Around $P_{\text{c}}$, $T_{\text{m1}}$ starts to increase rapidly and marge with $T^*$.
Then, a single peak remains in $\rho(T)$ and $T^*$ is renamed to $T_{\text{m2}}$, below which the Fermi liquid state is formed. \\
\item $T_{\text{m2}}$ increases with increasing $P$.
\end{enumerate}
These behaviors are actually reproduced by a theory including the CEF effect.\cite{Nishida06} 
However, the behavior of these characteristic temperatures in CeX$_{\text{c}}$ is slightly different:  
\begin{enumerate}
\item $T^*$ decreases with increasing $P$ much more rapidly at a rate of $-12.2$ K/GPa. \\
\item $T^*$ and $T_{\text{m2}}$ seems not to be connected in CeSe and in CeS. Double peak structure is still observed above $P_{\text{c}}$. 
\end{enumerate}

The slight decrease of $T^*$ with $P$ in general cases means that the true CEF level does not actually change with $P$. 
%The $P$ dependence of $T^*$ is explained theoretically by increasing the $c$-$f$ hybridization parameter $J_{cf}$ with keeping the CEF splitting unchanged. 
In the present cases of CeX$_{\text{c}}$, however, the rapid decrease of $T^*$ with $P$ seems to show that the actual CEF splitting decreases with $P$.\cite{Kawarasaki11,Takaguchi15} 
This is the most important and distinctive feature we consider in CeX$_{\text{c}}$, which is associated with the $f$-state degeneracy and the increase in $T_{\text{K}}$. 
This effect might also be associated with the anomalous increase of $S_{\text{L}}$ at high pressures. 

The critical pressure $P_{\text{c}}$ could also be affected by the pressure dependent CEF splitting. 
From Fig.~\ref{fig:2} we see that $P_{\text{c}}$ is the largest in CeSe and the lowest in CeTe. 
Although the $P$ dependence of $T_{\text{N}}$ of CeX$_{\text{c}}$ each follows that of the Doniach's diagram, 
this sequence of $P_{\text{c}}$ does not coincide with the that of the strength of $J_{cf}$, which is the smallest in CeTe and the largest in CeS. 
Here, it is necessary to take into account the difference in the CEF splitting, i.e., the largest in CeS and the smallest in CeTe. 
Especially in CeTe, the CEF splitting soon vanishes at around 2 GPa, where $T_{\text{N}}$ starts to decrease. 
The smallest $P_{\text{c}}$ in CeTe can be understood by considering the increased degeneracy of the $f$ state due to the collapsed CEF, which results in the increase of $T_{\text{K}}$ than that of CeSe and CeS. 
Then, $T_{\text{N}}$ of CeTe vanishes at the smallest pressure among the three compounds. 

We consider that the pressure dependence of $T_{\text{N}}$ in Fig.~\ref{fig:2} can be understood as a result of three effects. 
Firstly, with increasing $P$, $J_{cf}$ increases, which results in the increase in the RKKY interaction and $T_{\text{N}}$. 
Secondly, the $\Gamma_8$ CEF level falls down, which also contributes to the increase in $T_{\text{N}}$ because the $\Gamma_8$ state can give rise to larger magnetic moment. This is explained in Ref.~\citen{Takaguchi15} for CeTe by using a mean-field model calculation. 
These two effects both contribute to the increase in $T_{\text{N}}$. 
Thirdly, $T_{\text{K}}$ increases with $P$ through the increase of both $J_{cf}$ and the $f$-state degeneracy. 
In the present case of CeX$_{\text{c}}$, we suggest that the latter effect influences $T_{\text{K}}$ more than $J_{cf}$.

The MR effect can also be roughly explained. 
In CeSe and CeS below 2.5 GPa, the $\Gamma_8$ state is still well separated and the MR effect reflects the suppression of the Kondo effect within the $\Gamma_7$ ground state by the applied field. 
To understand the weak MR effects in CeSe and CeS shown in Fig.~\ref{fig:4}, Fig. 2 of Ref.~\citen{Kawakami86} based on a periodic Anderson model is a good reference. 
With increasing $P$, $T_{\text{K}}$ of the $\Gamma_7$ ground state increases, and the scaled parameters of $T/T_{\text{K}}$ and $H/T_{\text{K}}$ decrease. 
Since we are in the region of $T/T_{\text{K}}>1$, the increase of MR by the decrease of $T/T_{\text{K}}$ would be cancelled by the decrease of $H/T_{\text{K}}$. Then, the MR effect is expected to be weak in CeSe and CeS. 
To analyze in more detail to understand the larger $P$ dependence of MR in CeS than in CeSe, it is necessary to consider that $J_{cf}$ is larger in CeS than in CeSe (see $S_{\text{L}}$ in Fig.~\ref{fig:5}) and also that the pressure region of $0\sim 2.5$ GPa in CeS is more close to $P_{\text{c}}$, or the pressure where $S_{\text{L}}$ takes the maximum. 
It is also necessary to take into account the gradual increase of $f$ state degeneracy at high pressures.

The much larger MR effect in CeTe should be more related to the higher degeneracy due to the small CEF splitting, which almost vanishes above 1.5 GPa. Although $J_{cf}$ in CeTe is the smallest among the three compounds, the larger degeneracy of the $f$-state enhances the Kondo effect and leads to the large MR effect. 
It is remarked that, in Fig.~\ref{fig:4}, MR of CeTe increases with $P$ up to 1.5 GPa and becomes almost independent of $P$ above 1.5 GPa. 
This coincides with the shrinking process of CEF which is demonstrated by the variation of $T^*$ in Fig.~\ref{fig:5}(a). 
Above 1.5 GPa, the degeneracy of the ground state do not increase any more with $P$, and then the MR effect changes little with $P$, which may be understood in the same way as in CeSe and in CeS by the decrease of $T/T_{\text{K}}$ and $H/T_{\text{K}}$. 
Above 15 K, where the $\Gamma_8$ excited level is well populated, the $f$ state can be recognized as being fully degenerate, and the MR shows little $P$ dependence.

Finally, although we have studied a wide region of pressure, temperature, and magnetic field space, we remark that there still remain a low temperature region below 2 K that should be studied in future. 
As shown by the $\rho(T)$ curves in Fig.~\ref{fig:1}, the $\rho$ values at 2 K are still much higher than the residual resistivities, which are expected from the extrapolation of the $\rho(T)$ curves at 0 GPa. 
This shows that the system has not yet fallen to the ground state, especially near pressures around $P_{\text{c}}$, 
and there remain a possibility that some drastic phenomena occur below 2 K. 

\section{Conclusion}
We have studied the Kondo effect in CeS, CeSe, and CeTe by the electrical resistivity measurements under high pressures up to 8 GPa. 
The $\ln T$ term in the temperature dependence of the resistivity commonly increases with increasing pressure, indicating that the $c$-$f$ hybridization increases with pressure. 
There are two $\ln T$ regions. One is at high temperatures, where both $\Gamma_7$ and $\Gamma_8$ states are involved, and the other is at low temperatures, where the $\Gamma_7$ ground state is mainly involved, resulting in the peak structure in $\rho(T)$ around $T^*$ reflecting the CEF splitting. 
From the $\ln T$ term in the low temperature region, we see that $J_{cf}$ is the largest in CeS and the smallest in CeTe. 

CeX$_{\text{c}}$ under high pressure starts from the magnetic ordering regime and the pressure dependence of $T_{\text{N}}$ can basically be understood from the Doniach's diagram. 
However, it is necessary to take into account the significant decrease and collapse of the CEF splitting under pressure to interpret the sequence of the critical pressure, where $T_{\text{N}}$ vanishes; $P_{\text{c}}$(CeTe) $<P_{\text{c}}$(CeS) $<P_{\text{c}}$(CeSe), which contradicts with the sequence of $J_{cf}$. 
We suggest that the increase in $T_{\text{K}}$ under high pressure, which is brought about more by the increase of the $f$-state degeneracy than by the increase of $J_{cf}$, is responsible for the pressure dependence of $T_{\text{N}}$ and magnetoresistance. 

Finally, the appearance of the double peak structure in $\rho(T)$ in CeSe and CeS at high pressures near and above $P_{\text{c}}$ is also a distinctive feature in CeX$_{\text{c}}$. 
Since $T^*$ seems to be connected to the low-temperature peak at $T_{\text{m1}}$, the second peak at $T_{\text{m2}}$ may suggest an appearance of another energy scale associated with the formation of the Kondo singlet state. In any case, however, this is an open question to be studied in future.

\section*{Acknowledgements}
This work was carried out by the joint research in the Institute for Solid State Physics, the University of Tokyo, 
and was supported by JSPS KAKENHI Grant Numbers 2430087 and 15K05175.

\end{document}